\documentclass[aps,pra,twocolumn,longbibliography,superscriptaddress]{revtex4-2}									
\usepackage{amsmath}
\usepackage{amsthm}
\usepackage{latexsym}
\usepackage{amssymb}
\usepackage{bm}
\usepackage{bbm}
\usepackage{appendix}
\usepackage{graphics,epstopdf}
\usepackage{physics}
\usepackage{color}
\usepackage[colorlinks=true,linkcolor=blue,citecolor=green,plainpages=false,pdfpagelabels]{hyperref}
\usepackage[normalem]{ulem}
\usepackage{newlfont}
\usepackage{amsfonts}
\usepackage{amsthm}
\usepackage{graphicx}
\usepackage{epsfig}
\usepackage{mathrsfs}
\usepackage[thinc]{esdiff}
\usepackage{caption}
\usepackage{hyperref}
\usepackage{subcaption}
\newcommand\scalemath[2]{\scalebox{#1}{\mbox{\ensuremath{\displaystyle #2}}}}
\usepackage{times}

\newtheorem{proposition}{Proposition}

\newtheorem{theorem}{Theorem}
\newtheorem{dfn}{Definition}
\newenvironment{prof}{\noindent \textbf{Proof: }\ignorespaces}{\hspace*{\fill}$\Box$\medskip}
\DeclareOldFontCommand{\rm}{\normalfont\rmfamily}{\mathrm}

\def\tr{\operatorname{tr}}

\begin{document}
	
\title{Monogamy and tradeoff relations for wave-particle duality information}
\author{Shailja Kapoor}
\affiliation{Harish-Chandra Research Institute,  \\ A CI of Homi Bhabha National Institute, Chhatnag Road, Jhunsi, Allahabad 211  019, India }

\author{Sohail}
\affiliation{Harish-Chandra Research Institute,  \\ A CI of Homi Bhabha National Institute, Chhatnag Road, Jhunsi, Allahabad 211  019, India }

\author{Gautam Sharma}
\affiliation{QpiAI India Pvt. Ltd., Hub 1 SEZ Tower, Karle Town Centre, Nagavara, Bangalore 560045, India}
\author{Arun K. Pati}
\affiliation{Centre for Quantum Science and Technology,\\International Institute of Information Technology-Hyderabad, Gachibowli, Hyderabad, India.}
\begin{abstract}
The notions of predictability and visibility are essential in the mathematical formulation of wave particle duality. The work of Jakob and Bergou [Phys. Rev. A 76, 052107] generalises these notions for higher-dimensional quantum systems, which were initially defined for qubits, and subsequently proves a complementarity relation between predictability and visibility. By defining the single-party information content of a quantum system as the addition of predictability and visibility, and assuming that entanglement in a bipartite system in the form of concurrence mutually excludes the single-party information, the authors have proposed a complementarity relation between the concurrence and the single-party information content. We show that the information content of a quantum system defined by Jakob and Bergou is nothing but the Hilbert-Schmidt distance between the state of the quantum system of our consideration and the maximally mixed state. Motivated by the fact that the trace distance is a good measure of distance as compared to the Hilbert-Schmidt distance from the information theoretic point of view, we, in this work, define the information content of a quantum system as the trace distance between the quantum state and the maximally mixed state. We then employ the quantum Pinsker's inequality and the reverse Pinsker's inequality to derive a new complementarity and a reverse complementarity relation between the single-party information content and the entanglement present in a bipartite quantum system in a pure state. As a consequence of our findings, we show that for a bipartite system in a pure state, its entanglement and the predictabilities and visibilities associated with the subsystems cannot be arbitrarily small as well as arbitrarily large.
	
\end{abstract}

\maketitle
\section{Introduction}

The notion of complementarity was introduced by Niels Bohr\cite{BOHR1928} in order to explain the unusual behaviours of small particles such as electrons. Electrons in some experiments behave like waves, and in others they behave like particles, but they do not show these two different behaviours simultaneously. The wave nature and the particle nature of smaller objects like electrons are mutually exclusive. In the context of interference experiments, the particle nature is described by predictability, and the wave nature is described by visibility. In Young's double-slit experiment, predictability is captured by the difference in probability of the particle going through the two different slits. If the probabilities of the particle going through the slits are equal, then the predictability is zero, and when the difference is one, then we certainly know the slit through which the particle has gone. The visibility is captured by the contrast of the fringe pattern. The concept of complementarity is also studied experimentally as well as theoretically using a Mach-Zehnder interferometer, where the fringe visibility represents the wave nature, and the path predictability the particle nature of light. A quantitative description of wave-particle duality was first provided by the work of W. K. Wootters and W. H. Zurek \cite{Wooters_Zureh}. More specifically, if
\begin{align}
    \rho=\left(
\begin{array}{cc}
 \rho_{11} & \rho_{12} \\
 \rho_{21} & \rho_{22} \\
\end{array}
\right)
\end{align}
represents the quantum state of the qubit of our interest, then the predictability $P$ and the visibility $V$ is defined by the following relations:
\begin{align}
    P&=|\rho_{11}- \rho_{22}|, \label{predictability_qubit} \\
    V&=2|\rho_{12}|. \label{visibility_qubit}
\end{align}
The complementarity between the predictability and the visibility is mathematically captured by the the relation\cite{GREENBERGER1988391}:
\begin{align}
    P^2+ V^2 \leq 1.
\end{align}
The above complementarity relation has been verified in experiments with single photons, atoms, nuclear magnetic resonance, and faint lasers~\cite{PhysRevLett.81.5705, Xinhua_Peng_2003, PhysRevA.72.052109, PhysRevA.60.4285, PhysRevLett.100.220402,10.1119/1.2815364}.
The notion of wave particle-particle duality was generalised for composite systems in the context of which-way detection in Refs.~\cite{Jaeger_1, Jaeger_2, Englert}. There is an additional notion of complementarity in bipartite quantum systems. Roughly, it says that any single-partite information is mutually excluded by the amount of entanglement present in the bipartite quantum system~\cite{Jacob_2, Tessier2005, HOSOYA, Brukner2005, PhysRevA.71.062307_US}. The qualitative statement about the complementarity relation between entanglement and single-party information is made quantitative for a two-qubit system in a pure state by the authors in Ref~\cite{Jacob_2}. They have considered a pure state $\ket{\Psi}=a \ket{00}+ b \ket{01}+c \ket{10}+d \ket{11}$. The concurrence, which is a measure of entanglement, turns out to be $C(\ket{\Psi})=2 | ad-bc|$. The predictability and the visibility given by Eqs.~(\ref{predictability_qubit}) and (\ref{visibility_qubit}), respectively, turn out to be the following:

\begin{align}
\nonumber
    P_1 &=|(|c|^2 +|d|^2)-(|a|^2+|b|^2)|, \\
\nonumber    
    P_2 &= |(|b|^2 +|d|^2)-(|a|^2+|c|^2)|, \\
\nonumber    
    V_1 &= 2|ac^* + bd^*|, \\
\nonumber    
    V_1 &= 2|ab^* + cd^*|.
\end{align}
With theses equations, their derived complementarity relation is
\begin{align}
    [C(\ket{\Psi})]^2+ P^2_k+ V^2_k=1,
\end{align}
where $k=1,2$. Defining $\mathbb{S}^2_k = P^2_k+ V^2_k$, which the authors called the single party information content, the complementarity relation becomes:
\begin{align}
    [C(\ket{\Psi})]^2+ \mathbb{S}^2_k =1
\end{align}
Although the definitions of predictability and visibility have clear meaning in Young's double-slit type interferometers, in multiport interferometers, the meaning of predictability and visibility is not obvious despite some attempts. Some works in this direction can be found in the Refs.~\cite{PhysRevA.64.042113, PhysRevA.67.066101, Bimonte_2003, PhysRevA.64.042315, Luis_2001, PhysRevA.100.042105,QURESHI2017598,PhysRevA.92.012118,Quanta87,Qureshi:21,Roy_2019,PhysRevA.103.022219}. The main reason behind the difficulties is that the compact analytic form of visibility and predictability is not known. The authors of Refs.~\cite{PhysRevA.92.012118,QURESHI2017598} have quantified the wave nature in terms of quantum coherence and the particle nature by the upper bound of the success probability in the unambiguous quantum state discrimination~\cite{Helstrom1969, 1996quant.ph..1020F, doi:10.1080/00107510010002599, Janos_A_Bergou_2007, PhysRevLett.109.180501, PhysRevLett.113.020501, Bae_2015, PhysRevLett.80.4999, PhysRevA.64.062305,PhysRevA.69.050307, PhysRevA.72.012329}. Jakob and Bergou in Ref.~\cite{Jacob_1} have generalised the notion of predictability and visibility for a $n$-dimensional quantum state by representing it with a $(n^2 -1)$-dimensional Bloch vector using the generators of the SU(n) group \cite{PhysRev.70.460, PhysRevLett.47.838, PhysRevA.31.1299, PhysRevA.34.662, alicki2007quantum, mahler1995quantum, KIMURA2003339, PhysRevA.68.062322} and subsequently, the complementarity relation between the generalised predictability and the visibility has been derived. Further, based on the assumption that entanglement in a bipartite system in the form of concurrence mutually excludes single-partite information, the authors have proposed a complementarity relation between the generalised concurrence and the single-party information content.\\
What we observe is that the definition of the information content of a system introduced by Jakob and Bergou in Ref.~\cite{Jacob_1}, is nothing but the Hilbert-Schmidt distance between the state of the system and the maximally mixed state. However, from the information theoretic point of view, the Hilbert-Schmidt distance is not a ``good'' measure of distance. It lacks the property of contractivity under the action of a completely positive and trace-preserving maps (CPTP). In this work, we define the information content of a physical system as the trace distance between the state of the quantum system and the maximally mixed state on the same Hilbert space. The trace distance obeys various nice properties, including contractivity under the action of CPTP maps. We use quantum Pinsker's inequality to prove a different complementarity relation between the single-party information content and the amount of entanglement in a bipartite pure quantum state, with the entanglement being quantified by the von Neumann entropy of the reduced state. We also employ the reverse Pinsker's inequality to prove a reverse complementarity relation between the same. As a corollary of our findings, we provide a complementarity and a reverse complementarity relation between entanglement and single-partite generalised predictability and visibility.
\\
\\
The paper is organised as follows: In Section~(\ref{SEC_2}), we briefly describe the definitions of visibility and predictability as introduced in \cite{Jacob_1} and prove a monogamy relation between information contents across bi-partitions in a pure tripartite quantum state. In Section~(\ref{SEC_3}), we define the information content of a quantum system as the trace distance between the state of the system and the maximally mixed state. Then we employ the Pinsker's and the reverse Pinsker's inequality to prove a new complementarity and a reverse complementarity relation between the entanglement present in a pure bipartite quantum state and the single-party information contents. In Section~(\ref{SEC_4}), we conclude our paper.

\section{Complementarity relations for single and bipartite systems} \label{SEC_2}

In this section we briefly describe the definitions of visibility and predictability as introduced in \cite{Jacob_1}. These measures have been derived using generators of SU(n) group. For an $n$-dimensional system $\rho$  visibility and predictability are defined as 
\begin{align}
\mathcal{V}^2=2\sum^{n}_{j,k,j\neq k}|\rho_{jk}|^2. \label{generalised_visibility}
\end{align}

\begin{align}
\mathcal{P}^2=2\left(\sum_{j=1}^{n}\rho_{jj}^2-\frac{1}{n}\right). \label{generalised_predictability}
\end{align}

Upon addition, these quantities obey the following complementarity relation 
\begin{align}\label{singlecomp}
\mathbb{S}^2:=\mathcal{V}^2+\mathcal{P}^2=2\left(\mathrm{tr}(\rho^2)-\frac{1}{n}\right)\leq\frac{2(n-1)}{n}.
\end{align}

The quantity $\frac{2(n-1)}{n}$ on the right is the maximum possible length of the Bloch vector. Since the maximum length is achieved for pure states, the inequality is saturated if and only if $\rho$ is a pure state.  Both predictability and visibility are also bounded from above by the maximum length of the Bloch vector. The quantity $\mathbb{S}^2=\mathcal{P}^2+\mathcal{V}^2$ is considered as the total information content of the system and is upper bounded of the length of the Bloch vector which is the  intrinsic information in the system. Moreover, $\mathbb{S}$ is invariant under the action of unitary operators, whereas $\mathcal{P}$  and $\mathcal{V}$ vary with change in basis~\cite{PhysRevLett.83.3354, PhysRevA.63.022113, doi:10.1098/rsta.2001.0981, Brukner2003}. 
\\
In classical physics, the knowledge of the individual subsystem is enough to provide complete knowledge about the bipartite system. However, in quantum physics, the situation is completely different. To have complete knowledge about a bipartite quantum system, we need to look at the correlation between the subsystems. Entanglement in a composite system is one of such correlations. Apart from being a resource in various quantum information processing tasks~\cite{rep,Bennett,Bro1,Bro2,Steph,Jacob,Per,I1,I2,I3,D1,Agrawal,Sohail_tele}, it is a central object for the study of the foundations of quantum mechanics~\cite{PhysRev.47.777,b_a, b_r, b_e,b-1,b-2,b-3,b-4,PhysRevLett.128.160402}. Intuitively, the more the subsystems are correlated, the less information we get by looking at the individual subsystems. Thus, the entanglement of the composite system $\rho_{AB}$ affects the complementarity at the level of subsystems. For the subsystem $\rho_{k}$ with dimension $n_k$ and $k \in \{ A, B \}$, the following complementarity is proposed by the authors in Ref.~\cite{Jacob_1}:
\begin{align}\label{subsyscomp}
\mathcal{P}_k^2+\mathcal{V}_k^2+(\mathcal{C}_{AB}^n)^2=\mathbb{S}_k^2+(\mathcal{C}_{AB}^n)^2\leq \frac{2(n_k-1)}{n_k},
\end{align}
where $\mathcal{C}_{AB}^n$ is a proper generalisation of the concurrence for two qubits. The inequality is saturated if and only if $\rho_{AB}$ is a pure state. 
By adding the relation~(\ref{subsyscomp}) for $k=A$ and $k=B$, we have the following complementarity relation between the single party information contents of individual systems and the entanglement in the form of  generalised concurrence:
\begin{align}\label{bicomp}
\mathbb{S}_A^2+\mathbb{S}_B^2+2(\mathcal{C}_{AB}^n)^2  \leq \frac{2(n_A-1)}{n_A}+\frac{2(n_B-1)}{n_B},
\end{align}
where $n_A$ and $n_B$ are the dimensions of the respective subsystems. As the inequality in (\ref{subsyscomp}) saturates for pure bipartite state $\rho_{AB}$, the authors in Ref.~\cite{Jacob_1} have used it to define the generalized concurrence $\mathcal{C}_{AB}^n$ as $(\mathcal{C}_{AB}^n)^2:=2[1-\mathrm{tr}(\rho_{k}^2)]$, which is nothing but the linear entropy~\cite{principe2010information} of the reduced state $\rho_k$. Although linear entropy is used as a measure of entanglement in literature~\cite{Maleki:19, PhysRevA.84.042114,PhysRevA.75.032301, PhysRevA.73.052312, PhysRevA.74.022314,PhysRevB.104.134201}, it lacks additivity under the tensor product~\cite{PhysRevE.62.4665}. The von Neumann entropy, on the other hand, has the additivity property under tensor product~\cite{nielsen2010quantum}. In Section~\ref{SEC_3}, we will use the von Neumann entropy of the reduced density matrices of a pure bipartite quantum state as a measure of entanglement and prove our new complementarity and reverse complementarity relations.
\\
\subsection{Monogamy of information content}
In the above discussion we have seen that the total information of a bipartite state $\rho_{AB}$ is closely related with the entanglement present in a bipartite system. Motivated by this, here we seek to find a monogamy relation for the total information content of the bipartite states $\rho_{AB}$  and $\rho_{AC}$, i.e. when $\rho_{A}$ shares correlation with both $\rho_B$ and $\rho_C$. We consider both the scenarios when $\rho_{ABC}$ is pure as well as mixed.
\begin{proposition}
    For a tripartite system in a mixed state, the bipartite information contents of the subsystem AB and AC satisfies the following monogamy relation:
\begin{align}
\mathbb{S}^2_{AB}+\mathbb{S}^2_{AC}\leq (n_A + n_B) \mathbb{S}^2_{ABC} \label{Tripartite_complementarity_1}
\end{align}
\end{proposition}
\begin{prof}
 For a tripartite mixed state $\rho_{ABC}$, the following relations hold~\cite{Dupuis2014}:
 \begin{align}
     \mathrm{tr}(\rho^2_{AB}) \leq n_C \mathrm{tr}(\rho^2_{ABC}) \label{mixed_ineq_1}\\
      \mathrm{tr}(\rho^2_{AC}) \leq n_B \mathrm{tr}(\rho^2_{ABC}) \label{mixed_ineq_2}
 \end{align}
Now,
\begin{align}
\nonumber
    \mathbb{S}^2_{AB} + \mathbb{S}^2_{AC} = 2 \bigg( \mathrm{tr}(\rho^2_{AB}) + \mathrm{tr}(\rho^2_{AC}) -\frac{1}{n_{AB}} -\frac{1}{n_{AC}} \bigg)
\end{align}
Using the inequalities~(\ref{mixed_ineq_1}) and (\ref{mixed_ineq_2}), we have
\begin{align}
\nonumber
    \mathbb{S}^2_{AB} + \mathbb{S}^2_{AC} &\leq 2 \bigg( (n_B + n_C)\mathrm{tr}(\rho^2_{ABC}) -\frac{1}{n_{AB}} -\frac{1}{n_{AC}} \bigg)\\
    &= (n_B + n_C) \mathbb{S}^2_{ABC}
\end{align}
This completes the proof.
\end{prof}
\\
Using the fact that $\mathrm{tr}(\rho^2_{AB})+\mathrm{tr}(\rho^2_{AC}) \leq 2$, one can improve the bound in the inequality~(\ref{Tripartite_complementarity_1}) for a pure tripartite state as follows: 
\begin{align}
    \mathbb{S}^2_{AB}+\mathbb{S}^2_{AC}\leq 2\mathbb{S}_{ABC}^2. \label{Weaker_monogamy_tripartite}
\end{align}
The following proposition improves the bound of the inequality~(\ref{Weaker_monogamy_tripartite}).
\begin{proposition}
    Let a tripartite quantum system ABC be in a pure state. Then the bipartite information contents of the subsystem AB and AC satisfies the following monogamy relation:
    \begin{align}\label{monogamepure}
\mathbb{S}^2_{AB}+\mathbb{S}^2_{AC}\leq\frac{2n_{ABC}-n_C-n_B}{n_{ABC}-1}\mathbb{S}_{ABC}^2,
\end{align}
where $n_{ABC}= n_A n_B n_C$ with $n_A$, $n_B$ and $n_C$ being the dimensions of the subsystem A, B and C, respectively.
\end{proposition}
\begin{prof}
As the tripartite system is in a pure state, we have
    \begin{align}
        \mathbb{S}^2_{ABC}= 2 \bigg( 1 - \frac{1}{n_{ABC}}\bigg)
    \end{align}
Now,   
\begin{align*}
	& \mathbb{S}^2_{AB}+\mathbb{S}^2_{AC}\\
  &= 2\left(\mathrm{tr}(\rho_{AB}^2)+\mathrm{tr}(\rho_{AC}^2)-\frac{1}{n_An_B}-\frac{1}{n_An_C}\right) \nonumber \\
	 &\leq 2\left(2-\frac{n_C}{n_An_Bn_C}-\frac{n_B}{n_An_Bn_C}\right) \nonumber \\
	 &=2\left(\frac{2n_{ABC}-n_C-n_B}{n_{ABC}-1}\left(\frac{n_{ABC}-1}{n_{ABC}}\right)\right) \nonumber \\
	 &=\frac{2n_{ABC}-n_C-n_B}{n_{ABC}-1}\mathbb{S}_{ABC}^2.
\end{align*}
This completes the proof.
\end{prof}
\\
Now, it is straightforward to show that $\frac{2n_{ABC}-n_C-n_B}{n_{ABC}-1}\leq 2$, thus making this inequality tighter than the one in (\ref{Weaker_monogamy_tripartite}). 
\\
\\
We now prove that the information content $\mathbb{S}(\rho)$ defined in Eq.~(\ref{singlecomp}) is actually the Hilbert-Schmidt distance~\cite{nielsen2010quantum} between $\rho$ and the maximally mixed state on the same Hilbert space. The Hilbert-Schmidt distance between two Hilbert-Schmidt class operators $A$ and $B$ on a Hilbert space $\mathcal{H}$ is given by
\begin{align}
    \norm{A-B}^2_2= \mathrm{tr} \Big( (A-B)^{\dagger} (A-B) \Big)
\end{align}
Note that in finite dimensions every linear operator is a Hilbert-Schimdt operator.
\begin{proposition} \label{information_content_hilbert_schmidt}
    The information content $\mathbb{S}(\rho)$ of a quantum system in a state $\rho$ is equal to $\sqrt{2}$ times the Hilbert-Schmidt distance between $\rho$ and the maximally mixed state $\frac{\mathbb{I}}{n}$, with $n$ being the dimension of the Hilbert space associated with the system of our consideration, i.e.,
    \begin{align}
        \mathbb{S}(\rho)= \sqrt{2}\norm{\rho -\frac{\mathbb{I}}{n}}_{2}
    \end{align}
\end{proposition}
\begin{prof}
    We have,
    \begin{align}
    \nonumber
        2\norm{\rho -\frac{\mathbb{I}}{n}}^{2}_{2}&= 2\mathrm{tr} \bigg[ \bigg( \rho -\frac{\mathbb{I}}{n}\bigg)^2 \bigg]\\
    \nonumber    
        &= 2\mathrm{tr}  \bigg( \rho^2 -2\frac{\rho}{n}+\frac{\mathbb{I}}{n^2}\bigg)\\
    \nonumber    
        &= 2 \bigg( \mathrm{tr} (\rho^2) -\frac{1}{n}\bigg)\\
    \nonumber    
        &= \mathbb{S}^2(\rho),
    \end{align}
    which completes the proof.
\end{prof}
\begin{proposition}
    The information content $\mathbb{S}_{AB}$ of the joint system AB and the information contents $\mathbb{S}_A$ and $\mathbb{S}_{B}$ of the subsystems A and B, respectively satisfies the following relations:
    \begin{align}
        \mathbb{S}^2_{AB} \geq \frac{1}{n_B} \mathbb{S}^2_{A}, \label{DD_11}\\
        \mathbb{S}^2_{AB} \geq \frac{1}{n_A} \mathbb{S}^2_{B}. \label{DD_22}
    \end{align}
\end{proposition}
\begin{prof}
    For a mixed bipartite state $\rho_{AB}$ and its reduced states $\rho_A$ and $\rho_B$, the relations~\cite{Dupuis2014}
    \begin{align}
        \frac{1}{n_B} \leq \frac{ \mathrm{tr}(\rho^2_{AB})  }{ \mathrm{tr}(\rho^2_{A})} \leq n_{A}, \label{AA_11}\\
         \frac{1}{n_A} \leq \frac{ \mathrm{tr}(\rho^2_{AB})  }{ \mathrm{tr}(\rho^2_{B})} \leq n_{B} \label{BB_11}
    \end{align}
    holds. Now, from the definition of $ \mathbb{S}^2_{AB}$, we have
    \begin{align}
    \nonumber
         \mathbb{S}^2_{AB} &= 2 \bigg( \mathrm{tr} (\rho^2_{AB}) -\frac{1}{n_{AB}}\bigg) \\
    \nonumber     
         & \geq 2 \bigg( \frac{\mathrm{tr}(\rho^2_{A})}{n_B} -\frac{1}{n_{A}n_{B}}\bigg) \\
    \nonumber     
         &= \frac{1}{n_B} \mathbb{S}^2_{A},
    \end{align}
    where in the second line we have used the relation ~(\ref{AA_11}). Similarly, with the help of the relation~(\ref{BB_11}), one can prove that $\mathbb{S}^2_{AB} \geq \frac{1}{n_A} \mathbb{S}^2_{B}$.
\end{prof}
\\
Physically, the information content of a joint system should always be greater than or equal to the information content of its subsystems. The relations (\ref{DD_11}) and (\ref{DD_22}) do not guarantee such property for the information content $\mathbb{S}_{AB}$. More specifically, as the information content $\mathbb{S}_{AB}$ is the Hilbert-Schmidt distance between the quantum state $\rho_{AB}$ and the maximally mixed state $\frac{\mathbb{I}}{n_{AB}}$, and the Hilbert-Schmidt distance is not decreasing under a CPTP map~\cite{OZAWA2000158}, one cannot guarantee such a property for the information content $\mathbb{S}_{AB}$. Hence, $\mathbb{S}(\rho)$ is not a good measure of the information content for a quantum system.

\section{TRACE DISTANCE AS INFORMATION CONTENT and Complementarity relations in bipartite systems of arbitrary finite dimensions} \label{SEC_3}
We have seen in Proposition~(\ref{information_content_hilbert_schmidt}) that the information content defined by Jakob and Bergou in Ref.~\cite{Jacob_1} is nothing but the Hilbert-Schmidt distance between the state of the physical system of our consideration and the maximally mixed state. The distance between two quantum states quantifies the distinguishability between them. From the information theoretic point of view, any physical transformation of the quantum states cannot increase their distinguishability. Mathematically, this is equivalent to saying that any distance function defined on the space of quantum states should be contractive under the action of CPTP maps. However, the Hilbert-Schmidt distance is not contractive under CPTP maps, which implies that the definition of information content provided by Jakob and Bergou is not a good measure of information content. The trace distance between quantum states is contractive under the action of a CPTP map \cite{nielsen2010quantum}. Hence, we define the information content of a quantum system as the trace distance between the state of the physical system and the maximally mixed state. The trace norm of a trace-class operator $A$ on a Hilbert space $\mathcal{H}$ is defined as $\norm{A}_1:= \mathrm{tr}(\sqrt{A^{\dagger} A})$. It can be proved that the trace norm of a linear operator is equal to the sum of the modulus of its eigenvalues. The trace distance between two trace-class operators $A$ and $B$ is given by $\norm{A-B}_1$. It should be noted that in finite dimensions, every linear operator is a trace-class operator.
\\
\\
Now we provide a new definition of the information content of a quantum state and prove a complementarity and a reverse complementarity relation between the entanglement of a bipartite pure state and the single-party information content.
\begin{dfn}
    Let $\rho$ be a quantum state and let $\mathbb{I}$ be the identity operator on the Hilbert space $\mathcal{H}$ with the dimension being $n$. We define the information content $I(\rho)$ of the state $\rho$ to be the trace distance between the maximally mixed state $\frac{\mathbb{I}}{n}$ and $\rho$, i.e.,
    \begin{align}
        I(\rho):=\norm{\rho -\frac{\mathbb{I}}{n}}_{1}
    \end{align}
\end{dfn}
\begin{proposition} \label{HRI_1}
    The two different definition of information content $\mathbb{S}(\rho)$ and $I(\rho)$ satisfies the following relation when $\rho$ is a pure state.
    \begin{align}
        \mathbb{S}^{2}(\rho)=I(\rho)=2\left(1-\frac{1}{n}\right)
    \end{align}
\end{proposition}
\begin{prof}
    The definition of the two information content is the following:
    \begin{align}
        \mathbb{S}^2(\rho)&:=2\left(\mathrm{tr}(\rho^2)-\frac{1}{n}\right), \label{A1}\\
        I(\rho)&:=\norm{\rho -\frac{\mathbb{I}}{n}}_{1}. \label{A2}
    \end{align}
    When $\rho$ is pure, we have $\mathrm{tr}(\rho^{2})=1$ and the eigenvalues of $\rho$ are $1$ and rest of the $n-1$ eigenvalues are $0$. Hence Eq.~(\ref{A1}) and (\ref{A2}) becomes:
    \begin{align}
    \nonumber
        \mathbb{S}(\rho)&=2\left(1-\frac{1}{n}\right),
    \end{align}
    and 
    \begin{align}
    \nonumber
         I(\rho)&=\norm{\rho -\frac{\mathbb{I}}{n}}_{1}\\
    \nonumber     
         &=\bigg({1-\frac{1}{n}}\bigg) + (n-1) \frac{1}{n} \\
    \nonumber     
         &= 2\bigg({1-\frac{1}{n}}\bigg).
    \end{align}
    This completes the proof.
\end{prof}
\begin{proposition}
    The information content $I(\rho)$ of a quantum state $\rho$ is upper bounded by:
    \begin{align}
        I(\rho) \leq 2\bigg({1-\frac{1}{n}}\bigg), \label{SS_1}
    \end{align}
    where $n$ is the dimension of the Hilbert space associated with the system of our consideration. The equality holds when $\rho$ is a pure state.
\end{proposition}
\begin{prof} From the definition of the information content $I(\rho)$, we have
    \begin{align}
    \nonumber
          I(\rho)&=\norm{\rho -\frac{\mathbb{I}}{n}}_{1}\\
    \nonumber      
          &=\norm{\sum_{i}p_i \ketbra{\psi_i}{\psi_i} -\sum_{i}p_i\frac{\mathbb{I}}{n}}_{1}\\
    \nonumber      
          &=\norm{\sum_{i}p_i \left( \ketbra{\psi_i}{\psi_i} -\frac{\mathbb{I}}{n}\right)}_{1}\\
    \nonumber      
          & \leq \sum_{i}p_i\norm{\ketbra{\psi_i}{\psi_i} -\frac{\mathbb{I}}{n}}_{1}\\
    \nonumber      
          &= \sum_{i}p_i  2\bigg({1-\frac{1}{n}}\bigg)\\
    \nonumber      
          &=2\bigg({1-\frac{1}{n}}\bigg),
    \end{align}
    where in the second line, we have written the quantum state $\rho$ as a convex sum of pure state with $p_i$ being the probabilities, in the first inequality we have used the sub-additivity of norm and in the second last line we have used Proposition~(\ref{HRI_1}).
    This completes the proof. 
\end{prof}
\\
We see from Eqs.~(\ref{singlecomp}) and (\ref{SS_1}), that the information content $\mathbb{S}(\rho)$ and the information content $I(\rho)$ are both upper bounded by $2 \left( 1 -\frac{1}{n}\right)$. One can easily prove that the trace distance between two quantum states is larger than the Hilbert-Schmidt distance. In particular, we have \cite{Coles}
\begin{align}
    \sqrt{2}\norm{\rho -\sigma}_2 \leq \norm{\rho -\sigma}_1 \leq 2 \sqrt{R(\rho, \sigma)} \hspace{0.1cm}\norm{\rho -\sigma}_2,
\end{align}
where $R(\rho, \sigma)=\frac{rank(\rho)+ rank(\sigma)}{rank(\rho) rank(\sigma)}$.
 Using the above relation, we have the following relation between the information contents $\mathbb{S}(\rho)$ and $I(\rho)$:
\begin{align}
    \mathbb{S}(\rho) \leq I(\rho) \leq \sqrt{2R\bigg(\rho, \frac{\mathbb{I}}{n} \bigg)} \hspace{0.1cm}\mathbb{S}(\rho) \label{XYZ_1}
\end{align}
Hence the bound in Eq.~(\ref{SS_1}) is tighter than the one provided in Eq.~(\ref{singlecomp}).
\\
\\
We now state and prove our new complementarity and reverse complementarity relations. For a bipartite system AB in a pure state $\ket{\Psi}$ with reduced states $\rho_{A}:= \mathrm{tr}_{B}(\ket{\Psi}\bra{\Psi})$ and $\rho_{B}:= \mathrm{tr}_{A}(\ket{\Psi}\bra{\Psi})$, we quantify its entanglement $E(\Psi)$ by the von Neumann entropy of its reduced states, i.e., 
\begin{align}
    E(\Psi)=H(\rho_{k}):=-\mathrm{tr}(\rho_k \mathrm{ln}(\rho_k)),
\end{align}
where $k \in \{A,B\}$. In our proof, we will use relative entropy which is defined between two quantum state $\rho$ and $\sigma$ as~\cite{Umegaki1962}
 \begin{equation}
    D(\rho\Vert\sigma) := 
       \begin{cases}
                \tr(\rho(\log{\rho}-\log{\sigma})) &  \text{if}\  \operatorname{supp}(\rho)\subseteq \operatorname{supp}(\sigma), \\ 
                +\infty & \text{otherwise},
       \end{cases}\label{equ:relative entropy}
 \end{equation}
where $``\mathrm{supp}"$ represents the support of a linear operator.
\begin{theorem}
     Let a bipartite system $AB$ be in a pure state $\ket{\Psi}$. Then the following complementarity relation between the entanglement present in $\ket{\Psi}$ and the single party information content $I(\rho_k)$ holds:
    \begin{align}
        E(\Psi)+ \frac{1}{2\mathrm{ln}2} {I(\rho_k)}^{2} \leq \mathrm{ln} ({n_{k}}) \label{XYZ_2}
    \end{align}
    where, $n_{k}$ is the dimension of the Hilbert space  $\mathcal{H}_{k}$ associated with the subsystem $k$ and $k \in \{A,B\}$.
\end{theorem}
\begin{prof}
The von-Neuman entropy of $\rho_k$ defined  by $H(\rho_k) = -\mathrm{tr}(\rho_k \mathrm{ln}(\rho_k))$ can also be expressed in terms of the relative entropy between  $\rho_{k}$   and  $\frac{\mathbb{I}}{n_{k}}$ as follows:
\begin{align}
    H(\rho_{k})= \mathrm{ln}({n_k})- D\bigg(\rho_{k} \Big| \Big | \frac{\mathbb{I}}{n_k}\bigg).
\end{align}
We use the quantum Pinsker's inequality\cite{watrous_2018}:
\begin{align}
     D(\rho || \sigma ) \geq \frac{1}{2\mathrm{ln}2} \norm{\rho - \sigma}^{2}_{1}. \label{Pinsker}
\end{align}
Choosing $\rho =\rho_{k}$ and $\sigma = \frac{\mathbb{I}}{n_k}$ in the Pinsker's inequality~(\ref{Pinsker}), we have
\begin{align}
\nonumber
    D\bigg(\rho_{k} \Big| \Big | \frac{\mathbb{I}}{n_k}\bigg) \geq \frac{1}{2\mathrm{ln}2} \norm{\rho_{k} - \frac{\mathbb{I}}{n_k}}^{2}_{1}.
\end{align}
Now, using the definition of information content of the state $\rho_k$, we have:
\begin{align}
\nonumber
     & D\bigg(\rho_{k} \Big| \Big | \frac{\mathbb{I}}{n_k}\bigg) \geq \frac{1}{2\mathrm{ln}2} (I(\rho_{k}))^{2}\\
\nonumber     
     \implies &- D \bigg(\rho_{k} \Big| \Big | \frac{\mathbb{I}}{n_k} \bigg) \leq  - \frac{1}{2\mathrm{ln}2} (I(\rho_{k}))^{2} \\
\nonumber     
     \implies & \mathrm{ln}({n_k}) - D \bigg(\rho_{k} \Big| \Big | \frac{\mathbb{I}}{n_k}\bigg) \leq \mathrm{ln}({n_k}) - \frac{1}{2\mathrm{ln}2} (I(\rho_{k}))^{2} \\
\nonumber     
     \implies & H(\rho_{k}) \leq \mathrm{ln} ({n_{k}}) - \frac{1}{2\mathrm{ln}2} (I(\rho_{k}))^{2} \\
\nonumber     
     \implies & E(\Psi) + \frac{1}{2\mathrm{ln}2} (I(\rho_{k}))^{2} \leq \mathrm{ln} ({n_{k}})
\end{align}
This completes the proof.
\end{prof}
\\
Using the first inequality of relation~(\ref{XYZ_1}) in the relation~(\ref{XYZ_2}), we have the following complementarity relation which is less tight than the relation~(\ref{XYZ_2}):
\begin{align}
    E(\Psi) + \frac{1}{2\mathrm{ln}2} (\mathbb{S}(\rho_{k}))^{2} \leq \mathrm{ln} ({n_{k}})
\end{align}
The above complementarity relation can be expressed in terms of the predictability and the visibility defined in Eqs.~(\ref{generalised_predictability}) and~(\ref{generalised_visibility}), respectively, as follows:
\begin{align}
    E(\Psi) + \frac{1}{2\mathrm{ln}2} (\mathcal{P}_k^2 + \mathcal{V}_k^2) \leq \mathrm{ln} ({n_{k}}) \label{P_V_E_complementarity}
\end{align}
The complementarity relation in Eq.~(\ref{P_V_E_complementarity}) says that for a bipartite system in a pure state, its entanglement and the predictability and the visibility associated with the subsystems cannot be arbitrarily large.
By adding the complementarity relations~(\ref{P_V_E_complementarity}) for $k=A$ and $k=B$, we have the following complementarity relation between the entanglement and the single party predictabilities and the visibilities:
 \begin{align}
      2 E(\Psi) + \frac{1}{2\mathrm{ln}2}[\mathcal{P}_A^2 + \mathcal{V}_A^2 + \mathcal{P}_B^2 + \mathcal{V}_B^2]\leq \mathrm{ln}({n}). \label{P_V_E_complementarity_1}
 \end{align}
where $n= n_{A} n_{B}$ is the dimension of the Hilbert space associated with the bipartite system AB. The Pinsker's inequality (\ref{Pinsker}) can be refined by adding terms in the power of $\norm{\rho -\sigma}_1^2$ with positive coefficients~\cite{10.1063/1.4871575,1201071}. Using the refined Pinsker's inequality one can improve the complementarity relation~(\ref{XYZ_2}) and as a consequence the complementarity relations (\ref{P_V_E_complementarity}) and~(\ref{P_V_E_complementarity_1}) can be improved further.
\\
\\
Now, we will use the reverse Pinsker's inequality for finite dimensions \cite{Vershynina} to prove a reverse complementarity relation between entanglement present in a pure bipartite state and the single party information content. The reverse Pinsker's inequality says that,
\begin{align}
    D(\rho || \sigma) \leq M(\rho, \sigma)\norm{\rho -\sigma}_1 , \label{reverse_Pinskar_inequality}
\end{align}
where $M(\rho, \sigma)=\lambda_{\rho} \frac{\mathrm{ln}(\alpha_\rho)- \mathrm{ln}(\alpha_\sigma)  }{\alpha_\rho- \alpha_\sigma}$ with $\lambda_\rho$ being the maximum eigenvalue of $\rho$, and $\alpha_\rho$ and $\alpha_\sigma$ being the minimum non-zero eigenvalue of $\rho$ and $\sigma$, respectively.\\
\begin{theorem}
     Let a bipartite system $AB$ be in a pure state $\ket{\Psi}$. Then the following reverse complementarity relation between the entanglement present in $\ket{\Psi}$ and the single party information content $I(\rho_k)$ holds:
    \begin{align}
        E(\Psi) + M \bigg(\rho_k, \frac{\mathbb{I}}{n_{k}}\bigg) I({\rho_{k}}) \geq \mathrm{ln}(n_{k}) \label{WXY_1}
    \end{align}
    where $n_{k}$ is the dimension of the Hilbert space $\mathcal{H}_{k}$ associated with the subsystem $k$ and $k \in \{A,B\}$.
\end{theorem}
\begin{prof}
    Putting $\sigma=\frac{\mathbb{I}}{n_k}$ and $\rho= \rho_k$ in the relation~(\ref{reverse_Pinskar_inequality}), we have
\begin{align}
\nonumber
     & D\Big({\rho_{k}} \Big| \Big| \frac{\mathbb{I}}{n_{k}}\Big) \leq  M \bigg(\rho_k, \frac{\mathbb{I}}{n_{k}}\bigg)\norm{{\rho_{k}} -\frac{\mathbb{I}}{n_{k}}}_1 \\
\nonumber 
     \implies & -D\Big({\rho_{k}} \Big| \Big| \frac{\mathbb{I}}{n_{k}}\Big) \geq - M \bigg(\rho_k, \frac{\mathbb{I}}{n_{k}}\bigg) I({\rho_{k}})\\
\nonumber      
     \implies & H({\rho_{k}}) \geq \mathrm{ln}(n_{k}) - M \bigg(\rho_k, \frac{\mathbb{I}}{n_{k}}\bigg) I({\rho_{k}})\\   
\nonumber      
     \implies & E(\Psi) + M \bigg(\rho_k, \frac{\mathbb{I}}{n_{k}}\bigg) I({\rho_{k}}) \geq \mathrm{ln}(n_{k})       
\end{align}
This completes the proof.
\end{prof}
\\
Using the second inequality of relation~(\ref{XYZ_1}) in the relation~(\ref{WXY_1}), we have the following complementarity relation which is less tight than the relation~(\ref{WXY_1}):
\begin{align}
    E(\Psi) + \scalemath{0.85}{M \bigg(\rho_k, \frac{\mathbb{I}}{n_{k}}\bigg) \sqrt{2R\bigg(\rho_k, \frac{\mathbb{I}}{n_k} \bigg)} }\hspace{0.1cm}\mathbb{S}(\rho_k) \geq \mathrm{ln}(n_{k})
\end{align}
The above complementarity relation can be expressed in terms of the predictability and the visibility defined in Eqs.~(\ref{generalised_predictability}) and~(\ref{generalised_visibility}), respectively, as follows:\begin{align}
    E(\Psi) + \scalemath{0.85}{M \bigg(\rho_k, \frac{\mathbb{I}}{n_{k}}\bigg)} \sqrt{\scalemath{0.85}{2R\bigg(\rho_k, \frac{\mathbb{I}}{n_k} \bigg)(\mathcal{P}_k^2 + \mathcal{V}_k^2)}} \geq \mathrm{ln}(n_{k}) \label{R_P_V_E_complementarity}
\end{align}
The complementarity relation~(\ref{R_P_V_E_complementarity}) implies that for a bipartite system in a pure state, the entanglement, and the predictability and the visibility associated with the subsystems cannot be arbitrarily small.
\begin{proposition} \label{monotonicity_of_information_content_1}
    Let $\mathrm{\Phi}: \mathcal{B}(\mathcal{H}_A) \rightarrow \mathcal{B}(\mathcal{H}_B)$ be a quantum channel (also known as CPTP map) that preserves the maximally mixed state, i.e., $\mathrm{\Phi}(\frac{\mathbb{I}}{n_A})=\frac{\mathbb{I}}{n_B}$, where $n_A$ and $n_B$ are the dimensions of the Hilbert spaces $\mathcal{H}_A$ and $\mathcal{H}_B$ respectively and $\mathcal{B(H)}$ represents the space of the bounded linear operators on the Hilbert space $\mathcal{H}$. Then the information content of any state $\rho \in \mathcal{B}(\mathcal{H}_A)$ decreases under the action of such channels, i.e.,
    \begin{align}
        I(\rho) \geq I(\mathrm{\Phi}(\rho))
    \end{align}
\end{proposition}
\begin{prof}
    It is given that $\mathrm{\Phi}(\frac{\mathbb{I}}{n_A})=\frac{\mathbb{I}}{n_B}$ and as the trace-distance decreases under the action of a quantum channel, we have
    \begin{align}
    \nonumber
        I(\rho)&=\norm{\rho -\frac{\mathbb{I}}{n_A}}_{1}\\
    \nonumber    
        & \geq \norm{\mathrm{\Phi}(\rho) -\mathrm{\Phi}\bigg(\frac{\mathbb{I}}{n_A}\bigg)}_{1} \\
    \nonumber
        & \geq \norm{\mathrm{\Phi}(\rho) -\frac{\mathbb{I}}{n_B}}_{1} \\
    \nonumber
        & \geq I(\mathrm{\Phi}(\rho)).
    \end{align}
    This completes the proof.
\end{prof}
\\
We note that the partial trace is a CPTP map that preserves the maximally mixed state. Choosing $\mathrm{\Phi}= \mathrm{tr}_{B}$, we have $I(\rho_{AB}) \geq I(\mathrm{tr}_{B}(\rho_{AB}))=I(\rho_{A})$ and and choosing $\mathrm{\Phi}= \mathrm{tr}_{A}$, we have $I(\rho_{AB}) \geq I(\rho_{B})$. These two inequalities imply that the information content of a composite system is always greater than that of its subsystems. Using a similar argument, a weaker monogamy relation $I(\rho_{AB})+ I(\rho_{AC}) \leq 2I(\rho_{ABC})$ follows for tripartite systems. 
 
\section{Conclusion} \label{SEC_4}
In this work, we have derived a tighter monogamy relation between the information contents defined by Jakob and Bergou in Ref.~\cite{Jacob_1} across the possible bi-partitions of a tripartite state in a pure state. Then, we have proved that the definition of the information content of a quantum system as introduced in Ref.~\cite{Jacob_1} is nothing but the Hilbert-Schmidt distance between the state of the quantum system and the maximally mixed state. The Hilbert-Schmidt distance, however, lacks an essential property that a distance measure in quantum information should satisfy, namely the contractivity under the action of a completely positive and trace-preserving map. With this observation, we have defined the information content of a quantum system as the trace distance between the state of the system under our consideration and the maximally mixed state. We have then employed quantum Pinsker's inequality and the reverse Pinsker's inequality to derive a new complementarity and a reverse complementarity relation between the single-party information content and the entanglement present in a bipartite quantum system in a pure state. As a consequence of our findings, we have shown that for a bipartite system in a pure state, its entanglement, and the predictabilities and visibilities associated with the subsystems cannot be arbitrarily small as well as arbitrarily large.
\\
\\
\section{Acknowledgments} We acknowledge useful discussions with Ujjwal Sen. SK acknowledges support from HRI Prayagraj during her visit from April 2022 to June 2022. 
\bibliographystyle{unsrt}
\bibliography{Monogamyref}

\begin{thebibliography}{10}

\bibitem{BOHR1928}
N.~Bohr.
\newblock The quantum postulate and the recent development of atomic theory1.
\newblock {\em Nature}, 121(3050):580--590, Apr 1928.

\bibitem{Wooters_Zureh}
William~K. Wootters and Wojciech~H. Zurek.
\newblock Complementarity in the double-slit experiment: Quantum nonseparability and a quantitative statement of bohr's principle.
\newblock {\em Phys. Rev. D}, 19:473--484, Jan 1979.

\bibitem{GREENBERGER1988391}
Daniel~M. Greenberger and Allaine Yasin.
\newblock Simultaneous wave and particle knowledge in a neutron interferometer.
\newblock {\em Physics Letters A}, 128(8):391--394, 1988.

\bibitem{PhysRevLett.81.5705}
S.~D\"urr, T.~Nonn, and G.~Rempe.
\newblock Fringe visibility and which-way information in an atom interferometer.
\newblock {\em Phys. Rev. Lett.}, 81:5705--5709, Dec 1998.

\bibitem{Xinhua_Peng_2003}
Xinhua Peng, Xiwen Zhu, Ximing Fang, Mang Feng, Maili Liu, and Kelin Gao.
\newblock An interferometric complementarity experiment in a bulk nuclear magnetic resonance ensemble.
\newblock {\em Journal of Physics A: Mathematical and General}, 36(10):2555, feb 2003.

\bibitem{PhysRevA.72.052109}
Xinhua Peng, Xiwen Zhu, Dieter Suter, Jiangfeng Du, Maili Liu, and Kelin Gao.
\newblock Quantification of complementarity in multiqubit systems.
\newblock {\em Phys. Rev. A}, 72:052109, Nov 2005.

\bibitem{PhysRevA.60.4285}
Peter D.~D. Schwindt, Paul~G. Kwiat, and Berthold-Georg Englert.
\newblock Quantitative wave-particle duality and nonerasing quantum erasure.
\newblock {\em Phys. Rev. A}, 60:4285--4290, Dec 1999.

\bibitem{PhysRevLett.100.220402}
Vincent Jacques, E~Wu, Fr\'ed\'eric Grosshans, Fran\ifmmode \mbox{\c{c}}\else~\c{c}\fi{}ois Treussart, Philippe Grangier, Alain Aspect, and Jean-Fran\ifmmode \mbox{\c{c}}\else~\c{c}\fi{}ois Roch.
\newblock Delayed-choice test of quantum complementarity with interfering single photons.
\newblock {\em Phys. Rev. Lett.}, 100:220402, Jun 2008.

\bibitem{10.1119/1.2815364}
T.~L. Dimitrova and A.~Weis.
\newblock {The wave-particle duality of light: A demonstration experiment}.
\newblock {\em American Journal of Physics}, 76(2):137--142, 02 2008.

\bibitem{Jaeger_1}
Gregg Jaeger, Michael~A. Horne, and Abner Shimony.
\newblock Complementarity of one-particle and two-particle interference.
\newblock {\em Phys. Rev. A}, 48:1023--1027, Aug 1993.

\bibitem{Jaeger_2}
Gregg Jaeger, Abner Shimony, and Lev Vaidman.
\newblock Two interferometric complementarities.
\newblock {\em Phys. Rev. A}, 51:54--67, Jan 1995.

\bibitem{Englert}
Berthold-Georg Englert.
\newblock Fringe visibility and which-way information: An inequality.
\newblock {\em Phys. Rev. Lett.}, 77:2154--2157, Sep 1996.

\bibitem{Jacob_2}
Matthias {Jakob} and Janos~A. {Bergou}.
\newblock {Quantitative complementarity relations in bipartite systems}.
\newblock {\em arXiv e-prints}, pages quant--ph/0302075, February 2003.

\bibitem{Tessier2005}
Tracey~E. Tessier.
\newblock Complementarity relations for multi-qubit systems.
\newblock {\em Foundations of Physics Letters}, 18(2):107--121, Apr 2005.

\bibitem{HOSOYA}
A.~HOSOYA, A.~CARLINI, and S.~OKANO.
\newblock Complementarity of entanglement and interference.
\newblock {\em International Journal of Modern Physics C}, 17(04):493--509, 2006.

\bibitem{Brukner2005}
{\v{C}}aslav Brukner, Markus Aspelmeyer, and Anton Zeilinger.
\newblock Complementarity and information in ``delayed-choice for entanglement swapping''.
\newblock {\em Foundations of Physics}, 35(11):1909--1919, Nov 2005.

\bibitem{PhysRevA.71.062307_US}
Micha\l{} Horodecki, Pawe\l{} Horodecki, Ryszard Horodecki, Jonathan Oppenheim, Aditi Sen(De), Ujjwal Sen, and Barbara Synak-Radtke.
\newblock Local versus nonlocal information in quantum-information theory: Formalism and phenomena.
\newblock {\em Phys. Rev. A}, 71:062307, Jun 2005.

\bibitem{PhysRevA.64.042113}
Stephan D\"urr.
\newblock Quantitative wave-particle duality in multibeam interferometers.
\newblock {\em Phys. Rev. A}, 64:042113, Sep 2001.

\bibitem{PhysRevA.67.066101}
G.~Bimonte and R.~Musto.
\newblock Comment on ``quantitative wave-particle duality in multibeam interferometers''.
\newblock {\em Phys. Rev. A}, 67:066101, Jun 2003.

\bibitem{Bimonte_2003}
G~Bimonte and R~Musto.
\newblock On interferometric duality in multibeam experiments.
\newblock {\em Journal of Physics A: Mathematical and General}, 36(45):11481, oct 2003.

\bibitem{PhysRevA.64.042315}
Pranaw Rungta, V.~Bu\ifmmode~\check{z}\else \v{z}\fi{}ek, Carlton~M. Caves, M.~Hillery, and G.~J. Milburn.
\newblock Universal state inversion and concurrence in arbitrary dimensions.
\newblock {\em Phys. Rev. A}, 64:042315, Sep 2001.

\bibitem{Luis_2001}
Alfredo Luis.
\newblock Complementarity in multiple beam interference.
\newblock {\em Journal of Physics A: Mathematical and General}, 34(41):8597, oct 2001.

\bibitem{PhysRevA.100.042105}
Tabish Qureshi.
\newblock Interference visibility and wave-particle duality in multipath interference.
\newblock {\em Phys. Rev. A}, 100:042105, Oct 2019.

\bibitem{QURESHI2017598}
Tabish Qureshi and Mohd~Asad Siddiqui.
\newblock Wave–particle duality in n-path interference.
\newblock {\em Annals of Physics}, 385:598--604, 2017.

\bibitem{PhysRevA.92.012118}
Manabendra~Nath Bera, Tabish Qureshi, Mohd~Asad Siddiqui, and Arun~Kumar Pati.
\newblock Duality of quantum coherence and path distinguishability.
\newblock {\em Phys. Rev. A}, 92:012118, Jul 2015.

\bibitem{Quanta87}
Tabish Qureshi.
\newblock Coherence, interference and visibility.
\newblock {\em Quanta}, 8(1):24--35, 2019.

\bibitem{Qureshi:21}
Tabish Qureshi.
\newblock Predictability, distinguishability, and entanglement.
\newblock {\em Opt. Lett.}, 46(3):492--495, Feb 2021.

\bibitem{Roy_2019}
Prabuddha Roy and Tabish Qureshi.
\newblock Path predictability and quantum coherence in multi-slit interference.
\newblock {\em Physica Scripta}, 94(9):095004, jul 2019.

\bibitem{PhysRevA.103.022219}
Mohd~Asad Siddiqui and Tabish Qureshi.
\newblock Multipath wave-particle duality with a path detector in a quantum superposition.
\newblock {\em Phys. Rev. A}, 103:022219, Feb 2021.

\bibitem{Helstrom1969}
Carl~W. Helstrom.
\newblock Quantum detection and estimation theory.
\newblock {\em Journal of Statistical Physics}, 1(2):231--252, Jun 1969.

\bibitem{1996quant.ph..1020F}
Christopher~A. {Fuchs}.
\newblock {Distinguishability and Accessible Information in Quantum Theory}.
\newblock {\em arXiv e-prints}, pages quant--ph/9601020, January 1996.

\bibitem{doi:10.1080/00107510010002599}
Anthony Chefles.
\newblock Quantum state discrimination.
\newblock {\em Contemporary Physics}, 41(6):401--424, 2000.

\bibitem{Janos_A_Bergou_2007}
János~A Bergou.
\newblock Quantum state discrimination and selected applications.
\newblock {\em Journal of Physics: Conference Series}, 84(1):012001, oct 2007.

\bibitem{PhysRevLett.109.180501}
Gerald Waldherr, Adetunmise~C. Dada, Philipp Neumann, Fedor Jelezko, Erika Andersson, and J\"org Wrachtrup.
\newblock Distinguishing between nonorthogonal quantum states of a single nuclear spin.
\newblock {\em Phys. Rev. Lett.}, 109:180501, Nov 2012.

\bibitem{PhysRevLett.113.020501}
Megan Agnew, Eliot Bolduc, Kevin~J. Resch, Sonja Franke-Arnold, and Jonathan Leach.
\newblock Discriminating single-photon states unambiguously in high dimensions.
\newblock {\em Phys. Rev. Lett.}, 113:020501, Jul 2014.

\bibitem{Bae_2015}
Joonwoo Bae and Leong-Chuan Kwek.
\newblock Quantum state discrimination and its applications.
\newblock {\em Journal of Physics A: Mathematical and Theoretical}, 48(8):083001, jan 2015.

\bibitem{PhysRevLett.80.4999}
Lu-Ming Duan and Guang-Can Guo.
\newblock Probabilistic cloning and identification of linearly independent quantum states.
\newblock {\em Phys. Rev. Lett.}, 80:4999--5002, Jun 1998.

\bibitem{PhysRevA.64.062305}
Anthony Chefles.
\newblock Unambiguous discrimination between linearly dependent states with multiple copies.
\newblock {\em Phys. Rev. A}, 64:062305, Nov 2001.

\bibitem{PhysRevA.69.050307}
Anthony Chefles.
\newblock Condition for unambiguous state discrimination using local operations and classical communication.
\newblock {\em Phys. Rev. A}, 69:050307, May 2004.

\bibitem{PhysRevA.72.012329}
A.~K. Pati, P.~Parashar, and P.~Agrawal.
\newblock Probabilistic superdense coding.
\newblock {\em Phys. Rev. A}, 72:012329, Jul 2005.

\bibitem{Jacob_1}
Matthias Jakob and J\'anos~A. Bergou.
\newblock Complementarity and entanglement in bipartite qudit systems.
\newblock {\em Phys. Rev. A}, 76:052107, Nov 2007.

\bibitem{PhysRev.70.460}
F.~Bloch.
\newblock Nuclear induction.
\newblock {\em Phys. Rev.}, 70:460--474, Oct 1946.

\bibitem{PhysRevLett.47.838}
F.~T. Hioe and J.~H. Eberly.
\newblock $n$-level coherence vector and higher conservation laws in quantum optics and quantum mechanics.
\newblock {\em Phys. Rev. Lett.}, 47:838--841, Sep 1981.

\bibitem{PhysRevA.31.1299}
J.~P\"ottinger and K.~Lendi.
\newblock Generalized bloch equations for decaying systems.
\newblock {\em Phys. Rev. A}, 31:1299--1309, Mar 1985.

\bibitem{PhysRevA.34.662}
K.~Lendi.
\newblock Entropy production in coherence-vector formulation for n-level systems.
\newblock {\em Phys. Rev. A}, 34:662--663, Jul 1986.

\bibitem{alicki2007quantum}
R.~Alicki and K.~Lendi.
\newblock {\em Quantum Dynamical Semigroups and Applications}.
\newblock Lecture Notes in Physics. Springer Berlin Heidelberg, 2007.

\bibitem{mahler1995quantum}
G.~Mahler and V.~Weberru{\ss}.
\newblock {\em Quantum Networks: Dynamics of Open Nanostructures}.
\newblock Springer Berlin Heidelberg, 1995.

\bibitem{KIMURA2003339}
Gen Kimura.
\newblock The bloch vector for n-level systems.
\newblock {\em Physics Letters A}, 314(5):339--349, 2003.

\bibitem{PhysRevA.68.062322}
Mark~S. Byrd and Navin Khaneja.
\newblock Characterization of the positivity of the density matrix in terms of the coherence vector representation.
\newblock {\em Phys. Rev. A}, 68:062322, Dec 2003.

\bibitem{PhysRevLett.83.3354}
{\v{C}}aslav Brukner and Anton Zeilinger.
\newblock Operationally invariant information in quantum measurements.
\newblock {\em Phys. Rev. Lett.}, 83:3354--3357, Oct 1999.

\bibitem{PhysRevA.63.022113}
{\v{C}}aslav Brukner and Anton Zeilinger.
\newblock Conceptual inadequacy of the shannon information in quantum measurements.
\newblock {\em Phys. Rev. A}, 63:022113, Jan 2001.

\bibitem{doi:10.1098/rsta.2001.0981}
A.~Howie, J.~E. Ffowcs~Williams, Caslav Brukner, and Anton Zeilinger.
\newblock Young's experiment and the finiteness of information.
\newblock {\em Philosophical Transactions of the Royal Society of London. Series A: Mathematical, Physical and Engineering Sciences}, 360(1794):1061--1069, 2002.

\bibitem{Brukner2003}
{\v{C}}aslav Brukner and Anton Zeilinger.
\newblock {\em Information and Fundamental Elements of the Structure of Quantum Theory}, pages 323--354.
\newblock Springer Berlin Heidelberg, Berlin, Heidelberg, 2003.

\bibitem{rep}
Dik Bouwmeester, Jian-Wei Pan, Klaus Mattle, Manfred Eibl, Harald Weinfurter, and Anton Zeilinger.
\newblock Experimental quantum teleportation.
\newblock {\em Nature}, 390(6660):575--579, Dec 1997.

\bibitem{Bennett}
Charles~H. Bennett, Gilles Brassard, Claude Cr\'epeau, Richard Jozsa, Asher Peres, and William~K. Wootters.
\newblock Teleporting an unknown quantum state via dual classical and einstein-podolsky-rosen channels.
\newblock {\em Phys. Rev. Lett.}, 70:1895--1899, Mar 1993.

\bibitem{Bro1}
Sourav Chatterjee, Sk~Sazim, and Indranil Chakrabarty.
\newblock Broadcasting of quantum correlations: Possibilities and impossibilities.
\newblock {\em Phys. Rev. A}, 93:042309, Apr 2016.

\bibitem{Bro2}
Aditya Jain, Indranil Chakrabarty, and Sourav Chatterjee.
\newblock Asymmetric broadcasting of quantum correlations.
\newblock {\em Phys. Rev. A}, 99:022315, Feb 2019.

\bibitem{Steph}
Stephanie Wehner, David Elkouss, and Ronald Hanson.
\newblock Quantum internet: A vision for the road ahead.
\newblock {\em Science}, 362(6412), October 2018.

\bibitem{Jacob}
Jacob~D. Biamonte, Mauro Faccin, and Manlio~De Domenico.
\newblock Complex networks from classical to quantum.
\newblock {\em Communications Physics}, 2:1--10, 2017.

\bibitem{Per}
S.~Perseguers, M.~Lewenstein, A.~Ac{\'i}n, and J.~I. Cirac.
\newblock Quantum random networks.
\newblock {\em Nature Physics}, 6(7):539--543, Jul 2010.

\bibitem{I1}
Ryszard Horodecki, Michał Horodecki, and Paweł Horodecki.
\newblock Teleportation, bell's inequalities and inseparability.
\newblock {\em Physics Letters A}, 222(1):21--25, 1996.

\bibitem{I2}
I.~Chakrabarty.
\newblock Teleportation via a mixture of a two qubit subsystem of a n-qubit w and ghz state.
\newblock {\em The European Physical Journal D}, 57(2):265--269, Apr 2010.

\bibitem{I3}
S~Adhikari, N~Ganguly, I~Chakrabarty, and B~S Choudhury.
\newblock Quantum cloning, bell's inequality and teleportation.
\newblock {\em Journal of Physics A: Mathematical and Theoretical}, 41(41):415302, sep 2008.

\bibitem{D1}
Nirman Ganguly, Satyabrata Adhikari, A.~S. Majumdar, and Jyotishman Chatterjee.
\newblock Entanglement witness operator for quantum teleportation.
\newblock {\em Phys. Rev. Lett.}, 107:270501, Dec 2011.

\bibitem{Agrawal}
Pankaj Agrawal and Arun~K. Pati.
\newblock Probabilistic quantum teleportation.
\newblock {\em Physics Letters A}, 305(1):12--17, 2002.

\bibitem{Sohail_tele}
Sohail, Arun~K. Pati, Vijeth Aradhya, Indranil Chakrabarty, and Subhasree Patro.
\newblock Teleportation of quantum coherence.
\newblock {\em Phys. Rev. A}, 108:042620, Oct 2023.

\bibitem{PhysRev.47.777}
A.~Einstein, B.~Podolsky, and N.~Rosen.
\newblock Can quantum-mechanical description of physical reality be considered complete?
\newblock {\em Phys. Rev.}, 47:777--780, May 1935.

\bibitem{b_a}
J.~S. Bell.
\newblock {\em Speakable and Unspeakable in Quantum Mechanics: Collected Papers on Quantum Philosophy}.
\newblock Cambridge University Press, 2 edition, 2004.

\bibitem{b_r}
N.~Brunner, D.~Cavalcanti, S.~Pironio, V.~Scarani, and S.~Wehner.
\newblock {B}ell nonlocality.
\newblock {\em Rev. Mod. Phys.}, 86:419, Apr 2014.

\bibitem{b_e}
R.~F. Werner.
\newblock Quantum states with {E}instein-{P}odolsky-{R}osen correlations admitting a hidden-variable model.
\newblock {\em Phys. Rev. A}, 40:4277, Oct 1989.

\bibitem{b-1}
S.~J. Freedman and J.~F. Clauser.
\newblock Experimental test of local hidden-variable theories.
\newblock {\em Phys. Rev. Lett.}, 28:938, Apr 1972.

\bibitem{b-2}
A.~Aspect, P.~Grangier, and G.~Roger.
\newblock Experimental tests of realistic local theories via {B}ell's theorem.
\newblock {\em Phys. Rev. Lett.}, 47:460, Aug 1981.

\bibitem{b-3}
A.~Aspect, P.~Grangier, and G.~Roger.
\newblock Experimental realization of {E}instein-{P}odolsky-{R}osen-{B}ohm gedankenexperiment: A new violation of {B}ell's inequalities.
\newblock {\em Phys. Rev. Lett.}, 49:91, Jul 1982.

\bibitem{b-4}
A.~Aspect, J.~Dalibard, and G.~Roger.
\newblock Experimental test of {B}ell's inequalities using time-varying analyzers.
\newblock {\em Phys. Rev. Lett.}, 49:1804, Dec 1982.

\bibitem{PhysRevLett.128.160402}
Guillaume Aubrun, Ludovico Lami, Carlos Palazuelos, and Martin Pl\'avala.
\newblock Entanglement and superposition are equivalent concepts in any physical theory.
\newblock {\em Phys. Rev. Lett.}, 128:160402, Apr 2022.

\bibitem{principe2010information}
J.C. Principe.
\newblock {\em Information Theoretic Learning: Renyi's Entropy and Kernel Perspectives}.
\newblock Information Science and Statistics. Springer New York, 2010.

\bibitem{Maleki:19}
Yusef Maleki and Aleksei~M. Zheltikov.
\newblock Linear entropy of multiqutrit nonorthogonal states.
\newblock {\em Opt. Express}, 27(6):8291--8307, Mar 2019.

\bibitem{PhysRevA.84.042114}
Laura E.~C. Rosales-Z\'arate and P.~D. Drummond.
\newblock Linear entropy in quantum phase space.
\newblock {\em Phys. Rev. A}, 84:042114, Oct 2011.

\bibitem{PhysRevA.75.032301}
Fabrizio Buscemi, Paolo Bordone, and Andrea Bertoni.
\newblock Linear entropy as an entanglement measure in two-fermion systems.
\newblock {\em Phys. Rev. A}, 75:032301, Mar 2007.

\bibitem{PhysRevA.73.052312}
F.~Buscemi, P.~Bordone, and A.~Bertoni.
\newblock Entanglement dynamics of electron-electron scattering in low-dimensional semiconductor systems.
\newblock {\em Phys. Rev. A}, 73:052312, May 2006.

\bibitem{PhysRevA.74.022314}
Gustavo Rigolin, Thiago~R. de~Oliveira, and Marcos~C. de~Oliveira.
\newblock Operational classification and quantification of multipartite entangled states.
\newblock {\em Phys. Rev. A}, 74:022314, Aug 2006.

\bibitem{PhysRevB.104.134201}
G.~A. Canella and V.~V. Fran\ifmmode~\mbox{\c{c}}\else \c{c}\fi{}a.
\newblock Mott-anderson metal-insulator transitions from entanglement.
\newblock {\em Phys. Rev. B}, 104:134201, Oct 2021.

\bibitem{PhysRevE.62.4665}
G.~Manfredi and M.~R. Feix.
\newblock Entropy and wigner functions.
\newblock {\em Phys. Rev. E}, 62:4665--4674, Oct 2000.

\bibitem{nielsen2010quantum}
M.A. Nielsen and I.L. Chuang.
\newblock {\em Quantum Computation and Quantum Information: 10th Anniversary Edition}.
\newblock Cambridge University Press, 2010.

\bibitem{Dupuis2014}
Fr{\'e}d{\'e}ric Dupuis, Mario Berta, J{\"u}rg Wullschleger, and Renato Renner.
\newblock One-shot decoupling.
\newblock {\em Communications in Mathematical Physics}, 328(1):251--284, May 2014.

\bibitem{OZAWA2000158}
Masanao Ozawa.
\newblock Entanglement measures and the hilbert–schmidt distance.
\newblock {\em Physics Letters A}, 268(3):158--160, 2000.

\bibitem{Coles}
Patrick~J. Coles, M.~Cerezo, and Lukasz Cincio.
\newblock Strong bound between trace distance and hilbert-schmidt distance for low-rank states.
\newblock {\em Phys. Rev. A}, 100:022103, Aug 2019.

\bibitem{Umegaki1962}
Hisaharu Umegaki.
\newblock {Conditional expectation in an operator algebra. IV. Entropy and information}.
\newblock {\em Kodai Mathematical Seminar Reports}, 14(2):59 -- 85, 1962.

\bibitem{watrous_2018}
John Watrous.
\newblock {\em The Theory of Quantum Information}.
\newblock Cambridge University Press, 2018.

\bibitem{10.1063/1.4871575}
Eric~A. Carlen and Elliott~H. Lieb.
\newblock {Remainder terms for some quantum entropy inequalities}.
\newblock {\em Journal of Mathematical Physics}, 55(4):042201, 04 2014.

\bibitem{1201071}
A.A. Fedotov, P.~Harremoes, and F.~Topsoe.
\newblock Refinements of pinsker's inequality.
\newblock {\em IEEE Transactions on Information Theory}, 49(6):1491--1498, 2003.

\bibitem{Vershynina}
Anna Vershynina.
\newblock Quasi-relative entropy: the closest separable state and reversed pinsker inequality, 2021.

\end{thebibliography}
	
\end{document}